\documentclass[preprint,12pt]{elsarticle}

\usepackage{graphicx}
\usepackage{amssymb}
\usepackage{amsmath}
\usepackage{amsthm}

\usepackage{lineno}

\usepackage[usenames,dvipsnames]{color}

\journal{Journal Name}

\begin{document}

\begin{frontmatter}


\title{Dynamical reweighting methods for Markov models}



\author[post,equal]{Stefanie Kieninger}
\author[post,equal]{Luca Donati}
\author[post,email]{Bettina G.~Keller}

\address[post]{Freie Universit\"at Berlin, Department of Biology, Chemistry, Pharmacy, Arnimallee 22, D-14194 Berlin}
\address[equal]{equal contribiution}
\address[email]{bettina.keller@fu-berlin.de}

\begin{abstract}
Conformational dynamics is essential to biomolecular processes.
Markov State Models (MSM) are widely used to elucidate dynamic properties of molecular systems from unbiased Molecular Dynamics (MD).
However, the implementation of reweighting schemes for MSMs to analyze biased simulations is still at an early stage of development.
Several dynamical reweighing approaches have been proposed, which can be classified as approaches based on (i) Kramers rate theory,  (ii) rescaling of the probability density flux, (iii) reweighting by formulating a likelihood function, (iv) path reweighting. 
We present the state-of-the-art and discuss the methodological differences of these methods, their limitations and recent applications.
\end{abstract}  

\begin{keyword}
molecular simulation, Markov state models, dynamical reweighting, enhanced sampling, path reweighting, likelihood maximization, 
flux reweighting


\end{keyword}

\end{frontmatter}


%
\section{Introduction}
Biomolecular processes, such as ligand binding, protein-protein binding, protein folding, or allosteric changes, are brought about by conformational transitions. 
Within a single protein, different conformational transitions can occur on timescales ranging from picoseconds to seconds, 
and the exact balance between the stability of various conformational states, and the hierarchy of transition timescales is  essential for protein function \cite{Bowman2011, Lewandowski2015, Shaw2018, Bigman2020}.   
This enormous complexity arises, because the range of possible conformations (i.e.~the molecular state space $\Omega$) is vast, 
and the potential energy function, which determines the relative stability of the conformations has many minima which are separated
by barriers of vastly different heights.
Transitions between long-lived conformations often occur via a non-trivial series of short-lived conformations. 
Thus, an accurate model of the conformational dynamics from which the stability of conformations, transition rates and transition paths between conformations can be deduced is crucial for understanding biomolecular systems and processes.

Markov State Models (MSMs) \cite{Schuette1999b,Swope2004,Buchete2008,Keller2010,Prinz2011, Wang2018b} constructed from molecular dynamics simulations 
are a powerful tool to obtain dynamic information, such as long-lived conformations, the barriers separating them, 
and relaxation timescales for the equilibration across the barriers. 
However, it is precisely these barriers that pose a limitation to an MSM analysis.
Many molecules are characterized by high free energy barriers, and transitions across these barriers are rare events 
which are difficult to sample in an unbiased simulation.
Enhanced sampling techniques, such as umbrella sampling \cite{Torrie1977,Kaestner2005} and metadynamics \cite{Huber1994,Laio2002,Barducci2008}, facilitate the exploration of the state space by adding a bias $U(x)$ to the potential energy function $\widetilde V(x)$, such that the simulation is carried out with the potential $V(x) = \widetilde V(x) + U(x)$.
A variety of reweighting methods exists that recover stationary properties like ensemble averages or free energies from the biased simulations. 
But because transition rates and transition probabilities depend in a non-trivial way on the potential energy function, 
methods to recover dynamic information from biased simulations are much more difficult to develop.
Nonetheless, in recent years a number of dynamical reweighting approaches to tackle this problem have been proposed. 
These methods are derived from different formulations of molecular transitions, 
where each formulation gives rise to a different set of assumptions for the resulting dynamical reweighting method.
Starting with methods that are based on Kramers rate theory and proceeding to methods that reweight multi-state MSMs, we will discuss the assumptions of each approach and how they are mirrored by the applications that have been reported so far. 
Throughout the article, we will denote properties that correspond to the unbiased potential energy function $\widetilde{V}(x)$ by super-scripting them with a tilde, e.g.~$\widetilde{k}_{AB}$, whereas properties that correspond to the potential energy function $V(x)$
at which the simulation is carried out are denoted without decoration, e.g. $k_{AB}$.
To keep this contribution concise, we will limit the discussion to dynamical reweighting methods for simulations with biased potentials. 
This means that we will have to omit methods that recover dynamic information from simulations at elevated temperatures \cite{Chodera:2011ju, Prinz:2011, Stelzl2017a}, which are however closely related to dynamical reweighting methods for biased potentials.
Likewise, we will not cover path sampling strategies in which the enhanced sampling is achieved by respawning the simulations at certain checkpoints. For this topic we would like to point the reader to two excellent recent reviews \cite{Zuckerman:2017, Chong:2017bv}.

\section{Reweighting Kramers rate theory}
We denote the high-dimensional molecular state space by $\Omega$. 
$x$ is a specific point in this space (i.e.~a specific conformation), and $x_t$ denotes time-series of conformations (e.g.~a MD trajectory).
If the dynamics $x_t \in \Omega$ of a system is well characterized by two long-lived conformations $A$ and $B$ along a single reaction coordinate $q(x)$, 
separated by a transition state $q^*(x)$, 
the transition rate $k_{AB}$ from state $A$ to state $B$ can be modelled by Kramers rate theory \cite{Chandler1978, Hanggi1990}
\begin{equation}
  k_{AB}= \omega_A  \kappa_A \cdot \frac{Z^*}{Z_A} \, .
\label{eq:TST}
\end{equation}
Here,  $\omega_A$ is a characteristic frequency of the system in conformation $A$, 
$\kappa_A$ is the transmission coefficient, which models the possibility that the system will revert back to $A$, 
$Z_A = \frac{1}{Z}\int_{A} \exp (-\beta V(x))\, \mathrm{d}x$ is the partition function of conformation $A$,
$Z^* = \frac{1}{Z}\int_{q(x) = q^*(x)} \exp (-\beta V(x))\, \mathrm{d}x$ is the partition function of the transition state, 
and $Z$ is the partition function over the entire state space $\Omega$.
Thus, reweighting $k_{AB}$ to $\widetilde{k}_{AB}$ requires reweighting factors for $\omega_A$, $\kappa_A$, $Z_A$, and $Z^*$.

Several methods to derive reweighting factors for eq.~\ref{eq:TST} have been proposed, 
e.g. by reweighting the diffusion constant \cite{deOliveira:2007} of the diffusive dynamics along $q$, 
or by reformulating eq.~\ref{eq:TST} for a scaled potential \cite{Frank:2016jg}.
An enhanced sampling scheme which is particular suited for systems with two-state dynamics is metadynamics \cite{Huber1994, Laio2002}, 
and a corresponding reweighting scheme has been proposed by P.~Tiwary and M.~Parrinello \cite{Tiwary:2013}. 
The method uses an infrequent metadynamics protocol, in which the Gaussian biasing potentials are only deposited within states $A$ and $B$, but not in the transition region.
This has the advantage that, in the transition region, $V(x) \approx \widetilde V(x)$, and thus $Z^* \approx \widetilde Z^*$.
Since no reweighting factor for the transition region is needed, 
the reweighting factor is proportional to the ratio of the partition functions of state $A$
\begin{eqnarray}
	\frac{\widetilde k_{AB}}{k_{AB}} 
	= \frac{Z_A}{\widetilde Z_A} 
	&\propto& \frac{\int_A \exp(-\beta  V(x)) \, \mathrm{d}x}{\int_A \exp(-\beta (V(x) - U(x))) \, \mathrm{d}x}	
	= \frac{1}{\langle \exp (\beta U(x))\rangle_{V}}
	\label{eq:rew_Kramer}
\end{eqnarray}
where the subscript $V$ indicates that the ensemble average has been calculated for the ensemble at $V(x)$.
This reweighting scheme has been used to investigate several protein-ligand unbinding processes.
Estimates of $k_{\mathrm{off}}$ of an inhibitor from human p38 MAP kinase were in good agreement with the experimental value \cite{Casasnovas:2017}.
The reliability of the method was evaluated in a study of the dissociation process of two small different fragments from the protein FKBP by direct comparison to unbiased simulations \cite{Pramanik:2019}.
The unbinding process of a ligand from the L99A cavity mutant of T4 lysozyme has been studied using an 
improved infrequent metadynamics protocol
\cite{Wang:2018}.
Additionally, the combination of this reweighting scheme  with the variationally enhanced sampling \cite{Valsson:2014, Valsson:2016} has been demonstrated on 
alanine dipeptide in vacuum and benzophenone in water \cite{McCarty:2015}.

\section{Markov state models}
The dynamics of most molecular systems is much more complex than a simple two-state dynamics, 
exhibiting a multitude of metastable states separated by free energy barriers of various heights. 
Dynamics of this complexity are well captured by Markov state models (MSMs) \cite{Schuette1999b,Swope2004,Buchete2008,Keller2010,Prinz2011, Wang2018b}.
These models are built on the assumption that the time series of the molecular dynamics $x_t \in \Omega$  
is Markovian, ergodic, and fulfills detailed balance.
The high-dimensional state space $\Omega$ is discretized into a set of $n$ disjoint states $S  = \lbrace S_1, S_2, \dots S_n \rbrace$, 
and the time series $x_t$ is projected onto these states yielding a Markov jump process $s_t \in S$.
Often, the discretization is defined in terms of a lower-dimensional space of reaction coordinates.
Note that the states do not necessarily correspond to biologically relevant conformations. They are a partitioning of $\Omega$ into small cells.
The transition probability $p_{ij}(\tau)$ from $S_i$ to $S_j$ within a lag time $\tau$ can be estimated by counting the observed transitions $c_{ij}(\tau)$ in a trajectory of length $T$, and normalizing with respect to the number of outgoing transitions from $S_i$, 
\begin{eqnarray}
    p_{ij}(\tau) &=& \frac{c_{ij}(\tau)}{\sum_{j=1}^n c_{ij}(\tau)} \,.
\label{eq:transitionProb}    
\end{eqnarray}
The eigenvectors of the $n \times n$-transition matrix $\mathbf{P}(\tau)$, whose elements are $p_{ij}(\tau)$, 
contain a wealth of information on the dynamics of the system, long-lived conformations and the hierarchy of the barriers between them.
Eq.~\ref{eq:transitionProb} can be derived by maximizing the likelihood of transition probabilities $p_{ij}(\tau)$ given the observed time series $s_t$.
Alternatively, one can derive it by projecting the underlying dynamic operator onto the discretized state space. 
In this case, the transition counts can be understood as a path ensemble average
\begin{eqnarray}
    c_{ij}(\tau) 
    &=& \frac{1}{N_{\tau}}\sum_{t=0}^{N_{\tau}-1} \chi_i(x_t)\chi_j(x_{t+\tau})
\label{eq:counts}    
\end{eqnarray}
with 
\begin{eqnarray}
    \chi_i(x) &=&   \begin{cases}
                    1   &\mbox{if }  x \in S_i\\
                    0   &\mbox{otherwise} \, .
                    \end{cases}
\label{eq:characteristicFct}                    
\end{eqnarray}
The intuition for eqs.~\ref{eq:counts} and \ref{eq:characteristicFct} is:
Split up the trajectory $x_t$ into $N_{\tau}$ paths of length $\tau$. 
The starting time of each path is denoted by $t$, and also serves as a path index.
The argument in the sum evaluates to one, if the path represents a transition from $S_i$ to $S_j$ and to zero otherwise.
Thus, the sum counts the paths that represent transitions from $S_i$ to $S_j$. 
MSMs can also be derived using transition rates $k_{ij}$ rather than transition probabilities $p_{ij}(\tau)$. 
The corresponding rate matrix $\mathbf{K}$ is related to the transition matrix $\mathbf{P}(\tau)$
by $\exp(\mathbf{K}\tau) = \mathbf{P}(\tau)$.
To derive an estimator for $k_{ij}$ consider the transition probability density $p(x, y;\tau) \mathrm{d}y$, 
i.e.~the conditional probability density of finding the system in   $y\mathrm{d}y$ (i.e.~an infinitesimally small region around conformation $y$) at time $t+\tau$, given that it started in conformation $x$ at time $t$.
Assume that the starting point $x$ is located in $S_i$. 
For $\tau = 0$, $p(x, y;\tau) \mathrm{d}y$ is a delta-function centered at $x$. 
As $\tau$ increases, $p(x, y;\tau) \mathrm{d}y$ spreads out over the neighboring cells, 
thus creating a density flux across the boundaries of $S_i$. 
Using Gauss's divergence theorem on this flux, one obtains the following approximation for the rate constant at potential $V(x)$
\begin{equation}
    k_{ij} = \hat \Phi \cdot \sqrt{\frac{\pi_j}{\pi_i}} \, ,
    \label{eq:sqra}
\end{equation}
where $\pi_i$ and $\pi_j$ are the Boltzmann weights of the states $S_i$ and $S_j$. 
$\hat \Phi$ represents the flux through the intersecting surface between the states $S_i$ and $S_j$ 
in the absence of any potential energy function, which is scaled by $\sqrt{\pi_j / \pi_i}$.

\begin{figure}[ht]
\includegraphics[width=1.0\textwidth]{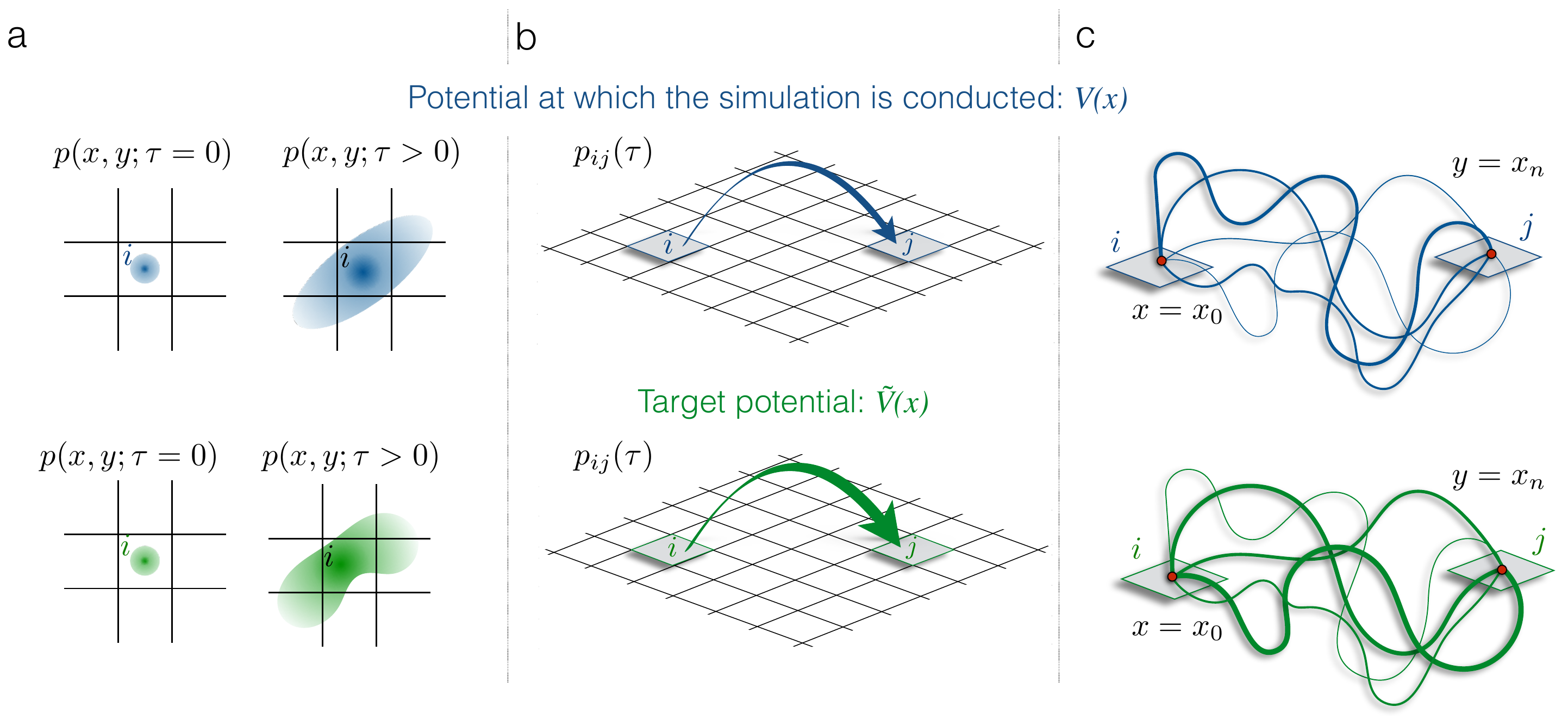}
\caption{Schematic representation of three different dynamical reweighting approaches to go from a biased dynamics (blue) to an unbiased dynamics (green). (a) Reweighting by rescaling the flux affects the flux through the intersecting surface between $S_i$ and $S_j$ created by the transition probability density $p(x,y;\tau)$. (b) Reweighting by formulating a likelihood function affects the transition probability $p_{ij}(\tau)$ from $S_i$ to $S_j$ within a lag time $\tau$. (c) In path reweighting, each path that starts in state $i$ and terminates in state $j$ has to be reweighted. Different probability values are represented by the thickness of the arrows and lines which connect two states.\label{fig:MSM}}
\end{figure}

This derivation of eq.~\ref{eq:sqra} via the transition probability density corresponds to discretizing the underlying dynamic operator 
(square-root approximation of the infinitesimal generator) \cite{Lie2013,Donati2018b, Heida2018}.
Eq.~\ref{eq:sqra} has originally been derived by D.~J.~Bicout and A.~Szabo by discretizing a one-dimensional Smoluchowski equation \cite{Bicout1998}.
It can also be obtained by exploiting the maximum caliber (maximum path entropy) approach \cite{Dixit2014, Dixit2015, Dixit2018, Stock2008, Otten2010}.
Eqs.~\ref{eq:transitionProb} to \ref{eq:sqra} show that there are three distinct ways to visualize a transition from $S_i$ to $S_j$.
The square-root approximation of the infinitesimal generator describes it in terms of a density flux through the boundaries of $S_i$
(Fig.~\ref{fig:MSM}.a).
The likelihood approach to MSMs visualizes the transition as a jump process, in which all information on what the system does between $t$
and $t +\tau$ is disregarded (Fig.~\ref{fig:MSM}.b).
Finally, in the path ensemble approach one imagines that a transition consists of a set of paths of length $\tau$, 
each of which starts in $S_i$ and ends in $S_j$. 
But some paths are more probable than others (Fig.~\ref{fig:MSM}.c).
Each of these images and the associated equations have given rise to a different dynamical reweighting approach.
In the following, we will discuss these approaches and their technical difficulties, 
starting with methods with multiple assumptions and going to methods with fewer assumptions.

\section{Reweighting by rescaling the flux} 
\label{sec:flux}
The starting point for this reweighting approach is eq.~\ref{eq:sqra}. 
In the derivation of this equation one has to make certain assumptions. 
The most important assumption is that the flux $\hat \Phi$ does not depend on the potential energy function.
Furthermore, the discretization $S$ of the state space into disjoint states has to satisfy the following requirements:
\begin{enumerate}
    \item The volumes of all states $S_i$ need to be approximately equal.
    \item The states $S_i$ need to be very small, such that the potential energy function in state $S_i$
          can be approximated by a constant:
          $V(x |x \in S_i) \approx V_i$.
\end{enumerate}
Thus, eq.~\ref{eq:sqra} and any reweighting method that is derived from it, requires an extremely fine discretization of the molecular state space. 
In practice, this means that the method can only be applied to systems whose dynamics can be described with only a few reaction coordinates that are dynamically well separated from the other degrees of freedom of the system. 
If this is the case, one can devise an appealingly simple reweighting strategy
\begin{eqnarray}
    \widetilde k_{ij} = \hat \Phi \sqrt{\frac{\widetilde \pi_j}{\widetilde \pi_i}}
    &\propto&       k_{ij} \cdot \sqrt{\frac{ \exp(\beta U_j)}{\exp(\beta U_i)}} 
    =			  k_{ij} \cdot \frac{1}{\exp\left( -\frac{U_j  - U_i}{2k_BT} \right)} \, ,
\label{eq:sqra_reweighting}
\end{eqnarray}
where we used that $\widetilde \pi_i \propto  \exp(-\beta (V_i - U_i))$.
The rates $\widetilde k_{ij}$ at the target potential $\widetilde V$ are obtained from the rates $k_{ij}$ at the potential $V(x)$, at which the simulation has been conducted,
by rescaling them with a factor that, besides temperature $T$, only depends on the potential in states $S_i$ and $S_j$. 
Note however that eq.~\ref{eq:sqra_reweighting} implicitly introduces two more conditions for this reweighting method:
\begin{enumerate}
\setcounter{enumi}{2}
    \item The simulation needs to be in local equilibrium within each state $S_i$ to yield a valid estimate of $k_{ij}$.
    \item The discretization $S$ needs to be chosen such that the discretization error of the MSM is small  
          at both potentials, $V(x)$ and $\widetilde V(x)$.
\end{enumerate}
In most cases the last requirement will not be critical, because conditions 1 and 2 already require a very fine discretization.

The reweighting factor in eq.~\ref{eq:sqra_reweighting} is used in the dynamic histogram analysis method (DHAM) \cite{Rosta:2014, Badaoui:2018ir}
to reweight transition probabilities 
\begin{equation}
    	\widetilde p_{ij}(\tau) = p_{ij}(\tau)\cdot \frac{1}{\exp\left( -\frac{U_j  - U_i}{2k_BT} \right)} \, .
\label{eq:DHAM}    	
\end{equation}
Applying the reweighting factor to transition probabilities rather than rates introduces a fifth condition:
\begin{enumerate}
\setcounter{enumi}{4}
    \item The lag time $\tau$ needs to be small.
\end{enumerate}
Note that DHAM has been derived using an MSM likelihood function. 
However the critical assumption in ref.~\cite{Rosta:2014} is that eq.~\ref{eq:sqra} holds.
DHAM has been used to study membrane permeabilities of a series of drug molecules \cite{Badaoui:2018ir} using umbrella sampling simulations. 
The dynamics was modeled along two reaction coordinates: 
the distance of the drug molecule to the center of the membrane, corresponding to the Cartesian $z$-coordinate of the system, 
and the projection of the orientation of the drug molecule onto this $z$-coordinate.
The DHAM analysis revealed complex free-energy landscapes with multiple minima and non-trivial pathways across the membrane.
From the DHAM-reweighted MSM, the authors calculated transition rates to enter, cross and exit the membrane, 
which were in close agreement with experiments and with results from long unbiased simulations. 
The system seems to be well suited for reweighting based on eq.~\ref{eq:sqra_reweighting} or eq.~\ref{eq:DHAM}, 
because membrane transition is slow compared to all other degrees of freedom in the system and can be described using only two reaction coordinates, 
which then have been discretized using an extremely fine grid with 1000 states. 

A similar reweighting scheme for MSMs has been derived from the principle of maximum caliber relative entropy, and has been successfully used to predict the effects of mutations on the folding kinetics of proteins \cite{Wan2016}.
Here, the first two time-independent components were used as reaction coordinates \cite{Schwantes2013,Hernandez2013}, and this two-dimensional space was discretized into 1000 states.

\section{Reweighting by formulating a likelihood function}
This reweighting approach considers simulations at a series of $K$ biasing potentials $V^{(k)}(x) = \widetilde V(x) + U^{(k)}(x)$, where $k$ denotes an index, 
as well as simulations at the unbiased potential $U^{(1)}(x) = 0$.
The likelihood of an MSM for the $k$th biasing potential is
\begin{eqnarray}
    \mathcal{L}_{\mathrm{MSM}}^{(k)} = \prod_{i=1}^n \prod_{j=1}^n \left(p_{ij}^{(k)}\right)^{c_{ij}^{(k)}}
\label{eq:likelihood}    
\end{eqnarray}
where $c_{ij}^{(k)}$ are the transition counts observed at potential $V^{(k)}(x)$.
$\mathcal{L}_{\mathrm{MSM}}^{(k)}$ can be maximized by varying $p_{ij}^{(k)}$ to obtain a stastically optimal MSM at potential $V^{(k)}(x)$, 
where constraints such as detailed balance or row-normalization of the transition matrix are incorporated by using Lagrange multipliers.
To reweight such a data set, one needs to combine the $K$ likelihood functions into an overall likelihood function, 
such that data from all potentials can be used to optimize the MSM at a specific $V^{(k)}(x)$. 
Except for the very specific conditions discussed in section \ref{sec:flux}, the transition probabilities at different potentials 
cannot be related to each other by closed-form reweighting factors. 
In general the MSMs are only linked via their equilibrium probabilities: 
$
\pi^{(k+1)}(x)/\pi^{(k)}(x) \propto \exp\left(-\beta U^{(k+1)}(x)\right)/ \exp\left(-\beta U^{(k)}(x) \right)
$.
The transition-based reweighting analysis method (TRAM) extends eq.~\ref{eq:likelihood} by a term that represents this connection between the $K$ different potentials \cite{Wu:2014, Wu:2016}.
The benefit of this approach is that the reweighted model only needs to fulfill the usual requirements of an MSM. 
Conditions 1, 2 and 5 in section \ref{sec:flux} can be dropped.
This benefit comes at a prize: 
To calculate the reweighted transition probabilities $\widetilde p_{ij}(\tau)$, one needs to numerically solve a system of coupled nonlinear equations
which arises from the maximization of the likelihood function.
Stelzl \emph{et.~al.} \cite{Stelzl:2017} derived a similar likelihood function starting from a rate matrix and incorporating the detailed balance conditions 
directly into the likelihood function. 
This yields the dynamic histogram analysis method extended to detailed balance (DHAMed). 
Ref.~\cite{Stelzl:2017} additionally contains a very detailed comparison of the TRAM and DHAMed equations.
Note that both methods require that the simulations at the target potential $\widetilde V(x)$ are included in the data set. 

TRAM has been used to study the complete binding equilibrium of a small inhibitor molecule, benzamidine, to the serine protease trypsin \cite{Wu:2016}.
The biased simulations were umbrella sampling simulations with umbrella potentials located along a one-dimensional reaction coordinate between the binding pose and a prebinding pose. 
The MSM was constructed on a two-dimensional space consisting of the umbrella sampling reaction coordinate and the second time-independent component \cite{Schwantes2013,Hernandez2013} of the system. 
This space was discretized into 100 states which highlights the fact 
that requirements 1 and 2 from section \ref{sec:flux} do not need to be fulfilled with this approach. 
The reweighted MSM revealed a complex binding mechanism including several secondary binding sites which acted as kinetic traps along the binding pathway.

TRAMMBAR \cite{Paul2017} is a variant of TRAM which allows for the construction of reweighted MSMs from Hamilton-replica exchange simulations \cite{Sugita1999}.
The method has been used to elucidate the full binding equilibrium of the 12-residue peptide inhibitor PMI to a large fragment of the oncoprotein Mdm2. 
The reweighted MSM showed a multivalent binding equilibrium in which the peptide binds in multiple different poses and contact surfaces to the protein, and the overall
stability of the complex is generated by the fact that these binding poses interconvert on various timescales without dissociation of the peptide from the protein.

%

\section{Path reweighting}
The starting point for the path-reweighting approach is eq.~\ref{eq:counts}, 
which interprets the counts as a sum over $N_{\tau}$ paths of length $\tau$. 
A path is the time series of positions in $\Omega$ generated by the MD simulation program
$(x_t, x_{t+\Delta t}, \dots, x_{t+ m \Delta t})$, where $\Delta t$ is the time step of the MD integrator, and $m \Delta t =\tau$.
Each path is generated with a certain frequency by the MD program 
that corresponds to its path probability $\pi(x_t)\cdot p(x_{t+\Delta t}, \dots, x_{t+ m \Delta t}| x_t)$ at the conditions of the simulations.
Thus, if the simulation is conducted at $V(x)$, each path enters the count estimate for $c_{ij}(\tau)$ with a weight of one, 
and the weighting factor does not need to be stated explicitly in eq.~\ref{eq:counts}.
However, a given path has a different probability $\widetilde \pi(x_t) \cdot \widetilde p(x_{t+\Delta t, \dots, x_{t+ m \Delta t}|x_t)}$ at $\widetilde V(x)$ 
than at $V(x)$. 
Therefore, when estimating the counts $\widetilde c_{ij}(\tau)$ at a potential $\widetilde V(x)$ from a set of paths simulated at $V(x)$, 
the weight with which each path enters the count estimate needs to be modified
\begin{eqnarray}
\widetilde{c}_{ij}(\tau) = \frac{1}{N_{\tau}} \sum\limits_{t=0}^{N_{\tau}-1} w(x_t, x_{t+\Delta t}, \dots, x_{t+ m \Delta t}) \, \chi_i(x_t)\chi_j(x_{t+\tau}) \, .
\end{eqnarray}
where $w(x_t, \dots, x_{t+ m \Delta t})  = (\widetilde{\pi}(x_t) \cdot \widetilde p(x_{t+\Delta t, \dots, x_{t+ m \Delta t}|x_t)}) / (\pi(x_t) \cdot p(x_{t+\Delta t}, \dots, x_{t+ m \Delta t}|x_t))$
is the ratio of the path probabilities.
The explicit expression of $w$ can be derived from the MD integrator. 
Usually, one uses an integrator for stochastic dynamics in which case the existence of $w$ is guaranteed by the Girsanov theorem \cite{Oksendal:2003}.
For overdamped Langevin dynamics, the expression for $w$ has been derived by L.~Onsager and M.~Machlup \cite{Onsager:1953}.
The path reweighting factor $w$ contains the gradient of the bias, $\nabla U(x)$, and is simple to calculate. 
However, it requires the knowledge of the molecular state at every integration time step. 
This is likely the reason that demonstrations of this reweighting approach have been limited to diffusion in low-dimensional energy landscapes
\cite{Woolf:1998, Zuckerman:1999kp, Zuckerman:2000gc, Schuette:2014} or short trajectories of alanine dipeptide \cite{Xing:2006},
in which writing the trajectory to a disc at every integration time step is still a viable option.
The problem can be solved by calculating the reweighting factor on-the-fly during the simulation \cite{Donati:2017}. 
This requires a modification of the MD program such that $\nabla U(x)$ is added to an internal variable at each integration time, and the variable is written to a file at the same frequency as the positions. 
Since in biased simulations, $\nabla U(x)$ is anyway calculated at each integration time step, this comes at a negligible computational cost. 
Ref.~\cite{Donati:2017} reports an implementation for calculating the path reweighting on-the-fly in OpenMM \cite{Eastman2013}.
It has been tested on reweighting MSM transition probabilities \cite{Donati:2017}, as well as mean first hitting times \cite{Quer:2017} in 
alanine dipeptide.
Similarly to the likelihood reweighting approach, requirements 1, 2 and 5 in section \ref{sec:flux} do not need to be fulfilled,
and the path reweighting method can thus be applied to MSMs constructed on high-dimensional reaction coordinate spaces. 
Additionally, since individual paths rather than transition probabilities are reweighted, condition 4 in section \ref{sec:flux} can be dropped. 
This can be a decisive advantage in systems with rugged potential energy surfaces, 
because the validity of an MSM critically depends on the discretization, which in turn is strongly influenced by the position and the height of the potential barriers. 
Thus, with path reweighting one can treat biases that alter the hierarchy of potential energy surface and even shift the position of barriers and optimize the discretization for the target potential $\widetilde V(x)$.
We are left with the requirement of local equilibrium (condition 3) which is still in place, 
because the relative probability for the initial position of each path enters the path reweighting factor $w$.

Path reweighting on-the-fly has been applied to reweight the folding equilibrium of a $\beta$-hairpin peptide from metadynamics simulations \cite{Donati:2018db}.
The metadynamics bias potential has been constructed along three reaction coordinates, which lead to a complete unfolding of the peptide. 
The path reweighting corrected for these fully unfolded structures, and revealed a dynamic equilibrium of partially unfolded structures. 
%
\section{Conclusion}
A wide collection of dynamical reweighting techniques has been compared, emphasizing the advancements and the limitations.
Tab.~\ref{tab:overview} summarizes the key points from the discussion.
The MD community has now a wide choice of methods which can be adapted to tackle different problems.
We believe that in the next decade, dynamical reweighting methods for MSMs will become fundamental to make new progresses on the knowledge of biomolecules, permitting a further development in drug design and in comprehension of human diseases.


\noindent
\begin{table}[h]
\begin{small}
\begin{tabular}{l | llll }
                            &Kramers    &&&\cr
                            &rate theory    &flux               &likelihood         &paths \cr
\hline  
number of long-lived  &&&&\cr              
conformations&2          &many           &many           &many   \cr        
\hline  
dimension of reaction           &&&&\cr
coordinate space                &1          &low                &high       &high \cr
\hline  
MSM discretization           &           &extremely  &&\cr           
$S$                                & -         &fine     &normal     &normal \cr
\hline  
optimizing $S$ for              &&&&\cr
$\widetilde V(x)$ possible                & -         & no                & no        &yes \cr
\hline  
technical difficulties          &           &                 & system of non-            &reweighting \cr
of reweighting                  & no        &no             &linear equations   &on-the-fly  \cr
\hline  
software   & implement            &implement                  &pyEMMA \cite{Scherer2015} &OpenMM \cite{Eastman2013} \cr
                & eq.~\ref{eq:rew_Kramer} &eq.~\ref{eq:sqra_reweighting} & for ref.~\cite{Wu:2014, Wu:2016}, &protocols in\cr 
                &                               &                                   &pyDHAMed for &supplement of\cr
                &                               &                                   &
                ref.~\cite{Stelzl:2017} & refs.~\cite{Donati:2017, Donati:2018db}
\end{tabular}
\caption{Requirements for successful dynamical reweighting via four different approaches, and available implementations}.
\label{tab:overview}
\end{small}
\end{table}

\section{Acknowledgments}
This research has been funded by Deutsche Forschungsgemeinschaft
(DFG) through grant CRC 1114 \emph{Scaling Cascades in Complex Systems}, Project B05 \emph{Origin of the scaling cascades in protein dynamics}.





\end{document}